\documentclass[twocolumn,showpacs,floats,floatfix,superscriptaddress,aps,pra]{revtex4-1}
\usepackage{amsfonts}
\usepackage{amssymb}
\usepackage{amsmath}
\usepackage{calc}
\usepackage{graphicx}
\usepackage{bm}

\usepackage[normalem]{ulem}
\usepackage{amsmath,amssymb}
\usepackage{multirow}
\usepackage{xcolor,soul}
\usepackage{srcltx}
\usepackage{hyperref,graphicx}

\def\be{ \begin{equation}}
\def\ee{ \end{equation}}
\def\bea{ \begin{eqnarray}}
\def\eea{ \end{eqnarray}}
\def\bse{ \begin{subequations}}
\def\ese{ \end{subequations}}
\def\bc{ \begin{center}}
\def\ec{ \end{center}}
\def\U{\mathbf{U}}
\def\H{\mathbf{H}}

\def\c{\mathbf{c}}

\def\M{\mathbf{M}}

\def\ket#1{\vert #1 \rangle}

\def\i{\text{i}}
\def\e{\text{e}}
\def\d{\text{d}}

\def\phase{\alpha}
\def\pha{\phi}
\def\phb{\xi}

\begin{document}

\author{Boyan Torosov}
\affiliation{Engineering Product Development, Singapore University of Technology and Design, 20 Dover Drive, 138682 Singapore}
\affiliation{Institute of Solid State Physics, Bulgarian Academy of Sciences, 72 Tsarigradsko chauss\'{e}e, 1784 Sofia, Bulgaria}
\author{Elica Kyoseva}
\affiliation{Engineering Product Development, Singapore University of Technology and Design, 20 Dover Drive, 138682 Singapore}
\author{Nikolay Vitanov}
\affiliation{Department of Physics, St Kliment Ohridski University of Sofia, 5 James Bourchier blvd, 1164 Sofia, Bulgaria}
\title{Fault-tolerant composite Householder reflection}


\begin{abstract}
We propose a fault-tolerant implementation of the quantum Householder reflection, which is a key operation in various quantum algorithms, quantum state engineering, generation of arbitrary unitaries, and entanglement characterization.
We construct this operation by using the modular approach of composite pulses and the relation between the Householder reflection and the quantum phase gate.
The proposed implementation is highly insensitive to variations in the experimental parameters, which makes it suitable for high-fidelity quantum information processing.
\end{abstract}

\maketitle

\section{Introduction}

The Householder reflection (HR) \cite{Householder} is a very powerful tool for a large variety of problems in data analysis: QR decomposition, least-square optimization, finding eigenvalues of large matrices, etc.
The decomposition of matrices by HR has been listed as one of the ten greatest discoveries in computational mathematics of 20th century by SIAM editors \cite{SIAM}.

Recently, it has been shown that HR has important applications in quantum physics and quantum information too, for instance in quantum algorithms \cite{Grover,SS}, for synthesis of unitary matrices \cite{IvanovKyoseva}, in quantum state engineering \cite{Ivanov2007}, as an entanglement witness \cite{witness}, etc.
Furthermore, it has been found \cite{Kyoseva2006,IvanovKyoseva} that HR is produced naturally by a particular quantum system: under certain conditions the propagator of an $N$-pod quantum system, consisting of $N$ degenerate states coupled simultaneously to another state, is given exactly by a HR.

To this end, the necessary conditions in the $N$-pod implementation are of resonant type (specific values of the interaction duration, the couplings and the detuning) and therefore, the thereby generated HR is prone to parameter errors, similarly to qubit addressing by resonant pulses of precise area.
In this paper we propose a fault-tolerant modular implementation of the HR operator by using the technique of composite pulses. The latter is a highly accurate and robust tool for quantum control, traditionally used in nuclear magnetic resonance (NMR) \cite{Levitt79,Freeman80,Levitt86,Freeman97}, quantum optics \cite{Torosov2011PRA,CAP,Genov2014PRL} and quantum computation \cite{Schmidt-Kaler,Timoney,Hill,Monroe}. Ideas similar to composite pulses have been used outside quantum physics too, e.g. in polarization optics \cite{polarization} and frequency conversion \cite{frequency conversion}.
A composite pulse is a sequence of pulses with well defined relative phases, which are used as control parameters and determined from the condition to produce a desired excitation profile.
Of special interest to the present problem are the broadband composite pulses, which produce high-fidelity excitation profiles, which are robust to variations in one or more experimental parameters around certain values of these parameters.
The construction of composite HR is based on a relation between the HR and the quantum phase gate, and uses the composite phase gates introduced recently \cite{PhaseGate}.

This paper is organized as follows.
In Sec. II we briefly sketch the theory behind composite pulses and introduce the HR.
In Sec. III we discuss how to combine these in order to obtain a robust composite HR.
Section IV presents the conclusions.

\section{Composite pulses and Householder reflections}

In this section we briefly review the theory of composite pulses, we show how they can be used to produce an error-resilient phase gate, and we introduce the Householder transformation.

\subsection{Composite pulses}

To explain the idea of CPs, let us consider a simple two-state quantum system, coherently driven by an external field.
Such a system is described by the time-dependent Schr\"{o}dinger equation
\be\label{Schr}
\i \hbar\partial_t \mathbf{c}(t) = \H(t)\mathbf{c}(t),
\ee
where $\c(t)=[c_1(t),c_2(t)]^T$ is a column vector containing the probability amplitudes of the two states $|\psi_1\rangle$ and $|\psi_2\rangle$, and the Hamiltonian is
\be
 \H(t) = (\hbar/2) \Omega(t) \e^{-\i D(t)} |\psi_1\rangle\langle \psi_2| + \text{h.c.}
\ee
Here $\Omega(t)$ is the Rabi frequency and $D(t) =\int_{0}^{t}\Delta(t^{\prime})\d t^{\prime}$,
 where $\Delta = \omega_0-\omega$ is the detuning between the field frequency $\omega$ and the Bohr transition frequency $\omega_0$.
 [Time dependence may be present in $\Delta$ due to time-dependent (chirped) field frequency or dynamic Stark or Zeeman shift in $\omega_0$.]
The propagator of the system, which is an operator that connects the initial and final amplitudes, $\c(t_f)=\U \c(t_i)$, can be parameterized by using the two complex Cayley-Klein parameters $a$ and $b$,
\be\label{U}
\U = \left[\begin{array}{cc}  a  & b \\  -b^{\ast} & a^{\ast} \end{array} \right].
\ee
A constant phase shift in the Rabi frequency $\Omega(t)\to\Omega(t)\e^{\i\phi}$ is imprinted onto the off-diagonal elements of the propagator,
\be\label{Uphi}
\U_{\phi} = \left[\begin{array}{cc}  a  & b \e^{\i\phi} \\  -b^{\ast}\e^{-\i\phi} & a^{\ast} \end{array} \right].
\ee
A composite pulse is by definition a sequence of pulses with different phases.
The propagator of a composite pulse, for a sequence of $n$ pulses, is
\be
\U^{(n)}=\U(\pha_n)\cdots\U(\pha_2)\U(\pha_1) .
\ee
If the phases $\pha_k$ are chosen appropriately, the propagator $\U^{(n)}$ can be made much more robust to variations in the experimental parameters than the single-pulse propagator $\U$.
This is the basic idea behind CPs and in such a way one can produce a huge variety of broadband (BB), narrowband (NB), and passband (PB) CPs with respect to variations in essentially any experimental parameter \cite{Torosov2011PRA}.
It is of particular relevance to us how CPs can be used to construct a quantum phase gate.

\subsection{Composite phase gate}

\begin{figure}[tb]
\includegraphics[width=5.5cm]{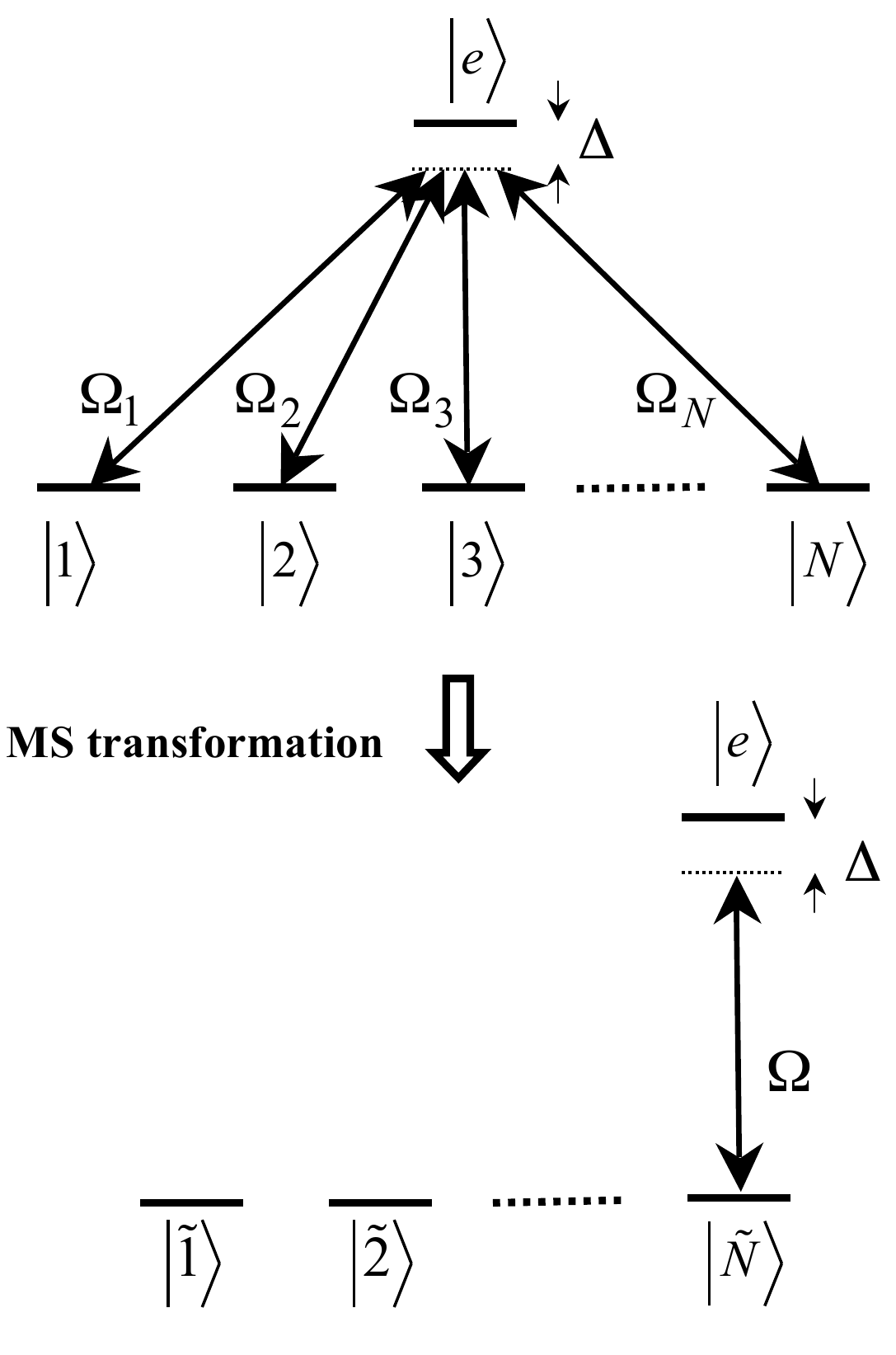}
\caption{(Top) Level scheme for the $N$-pod, which realizes the Householder transformation. (Bottom) The level scheme after the Morris-Shore transformation.}
\label{scheme}
\end{figure}

A single-qubit phase gate is defined as the $2\times 2$ matrix
\be\label{phase gate}
\Phi = \left[\begin{array}{cc}  e^{i\phase/2} & 0 \\  0 & e^{-i\phase/2} \end{array} \right],
\ee
where $\alpha$ is the phase difference between the two states of the qubit, accumulated due to the gate operation.
It was recently demonstrated \cite{PhaseGate}, that the propagator of a sequence of two CPs can be made equal to $\Phi$. Explicitly, we have
\be\label{phase gate propagator}
\Phi=\U_{\text{CP}_2}\U_{\text{CP}_1} ,
\ee
where
\bse\label{CP1&CP2}
\begin{align}
&\U_{\text{CP}_1}=\U(\pha_n)\cdots\U(\pha_2)\U(\pha_1), \\
&\U_{\text{CP}_2}=\U(\phb_n)\cdots\U(\phb_2)\U(\phb_1),
\end{align}
\ese
and
\be\label{phases second CP}
\phb_k = \pha_k+\pi+\phase/2.
\ee
The phases $\pha_k$ and $\phb_k$ are just the phases of the CPs and are chosen depending on which type of error-resilient CP we want to construct.
A detailed description of the composite phase gate is presented elsewhere \cite{PhaseGate}.

\subsection{Householder reflection}

The \emph{standard} HR is defined as
\begin{equation}\label{SHR}
\mathbf{M}(v)=\mathbf{I}-2\left\vert v\right\rangle \left\langle
v\right\vert ,
\end{equation}
where $\mathbf{I}$ is the identity operator and $\left\vert v\right\rangle$ is an $N$-dimensional normalized complex column-vector.
The HR \eqref{SHR} is both hermitian and unitary, $\mathbf{M}=\mathbf{M}^{^{\dagger }}=\mathbf{M}^{-1}$, which means that $\mathbf{M}$ is involutary, $\mathbf{M}^{2}=\mathbf{I}$. In addition, $\det \mathbf{M}=-1$. For real $\left\vert v\right\rangle $ the Householder transformation \eqref{SHR} has a simple
geometric interpretation: reflection with respect to an $(N-1)$-dimensional plane with a normal vector $\left\vert v\right\rangle $. In general, the vector $\left\vert v\right\rangle $ is complex and it is characterized by $2N-2$ real parameters (two parameters are discounted due to the normalization condition and the unimportant global phase).

The \emph{generalized} HR is defined as
\begin{equation}\label{GHR}
\mathbf{M}(v;\varphi )=\mathbf{I}+\left( e^{i\varphi }-1\right) \left\vert v\right\rangle \left\langle v\right\vert ,
\end{equation}
where $\varphi $ is an arbitrary phase. The standard HR \eqref{SHR} is a
special case of the generalized HR (\ref{GHR}) for $\varphi =\pi $: $%
\mathbf{M}(v;\pi )\equiv \mathbf{M}(v)$. The generalized QHR is unitary, $%
\mathbf{M}(v;\varphi )^{-1}=\mathbf{M}(v;\varphi )^{\dagger }=\mathbf{M}%
(v;-\varphi )$, and its determinant is $\det \mathbf{M}=e^{i\varphi }$.

It was shown earlier \cite{IvanovKyoseva,Kyoseva2006}, that the standard and generalized HR can be realized in an $N$-pod quantum system, wherein $N$ degenerate states are coupled to an ancillary state, as shown in Fig.~\ref{scheme}, or by using a similar coupling scheme and a STIRAP process \cite{Rousseaux2013}.
We will now briefly review the implementation of HR, and then show how it can be improved by CPs.
Let us assume that the couplings $\Omega_k$ in Fig.~\ref{scheme} have the same time dependence $f(t)$, but different amplitudes $\chi_k$ and phases $\beta_k$,
\be
\Omega_k(t)=\chi_k f(t) \e^{\i\beta_k} .
\ee
Such a system is described by the Hamiltonian
\be\label{HHR}
\H =\frac{\hbar}{2}
\left[\begin{array}{ccccc}
0 & 0 & \cdots & 0 & \Omega_1(t) \\
0 & 0 & \cdots & 0 & \Omega_2(t) \\
\vdots & \vdots & \ddots & \vdots & \vdots \\
0 & 0 & \cdots & 0 & \Omega_N(t) \\
\Omega_1^{\ast}(t) & \Omega_2^{\ast}(t) & \cdots & \Omega_N^{\ast}(t) & 2\Delta(t) \\
\end{array} \right].
\ee
By using the Morris-Shore transformation \cite{MS} one can reduce this $N+1$ state problem to a set of $N-1$ uncoupled states and a two-state system with a Hamiltonian, which involves the same detuning $\Delta$ as in Eq. \eqref{HHR} and the coupling is the root-mean-square (rms) Rabi frequency~\cite{note} $\Omega(t)=\sqrt{\sum_{k=1}^{N}|\Omega_k(t)|^2}$,
\be\label{MS-2SS}
\H_{\text{MS}} = \frac{\hbar}{2}
\left[\begin{array}{cc}
 0 & \Omega(t) \\
 \Omega(t) & 2\Delta(t)
\end{array} \right].
\ee
The propagator of this MS two-state system can be written as
\be\label{U_MS}
\U_{\text{MS}} = \left[\begin{array}{cc}  a  & b \\  -b^{\ast} & a^{\ast} \end{array} \right].
\ee
It is straightforward to verify that if $a=\e^{\i\varphi}$ (and hence, $b=0$) the propagator of the original $N$-state degenerate manifold is equal to the generalized HR \eqref{GHR},
\be
\U = \M(v,\varphi),
\ee
with
\be
v= \frac{1}{\chi}[\chi_1\e^{\i\beta_1},\chi_2\e^{\i\beta_2},\ldots,\chi_N\e^{\i\beta_N}]^{\text{T}},
\ee
where $\chi=\sqrt{\sum_{k=1}^{N}{\chi_k^2}}$ is the rms peak Rabi frequency and $\beta_{km}=\beta_{k}-\beta_{m}$.

The condition $a=\e^{\i\varphi}$, as seen from Eqs. \eqref{U} and \eqref{phase gate}, corresponds exactly to a phase gate in the MS two-state system \eqref{MS-2SS}, with $\alpha=2\varphi$.
Traditionally, there are several ways to produce such phase shifts. One way is to use a dynamic phase gate \cite{dynamic}, which only requires a single far-off-resonant pulsed field. Another basic approach is the geometric phase gate \cite{geometric}, which has certain advantages in terms of robustness against parameter fluctuations that come at the cost of more demanding implementations. An alternative phase gate uses adiabatic passage and relative laser phases \cite{laser phases}.
In the present work, we use the approach, based on composite pulses, which has been described in \cite{PhaseGate}, and we apply the same approach to construct robust and high-fidelity HRs.


\section{Composite Householder reflection}

\begin{figure}[t]
\includegraphics[width=8.5cm]{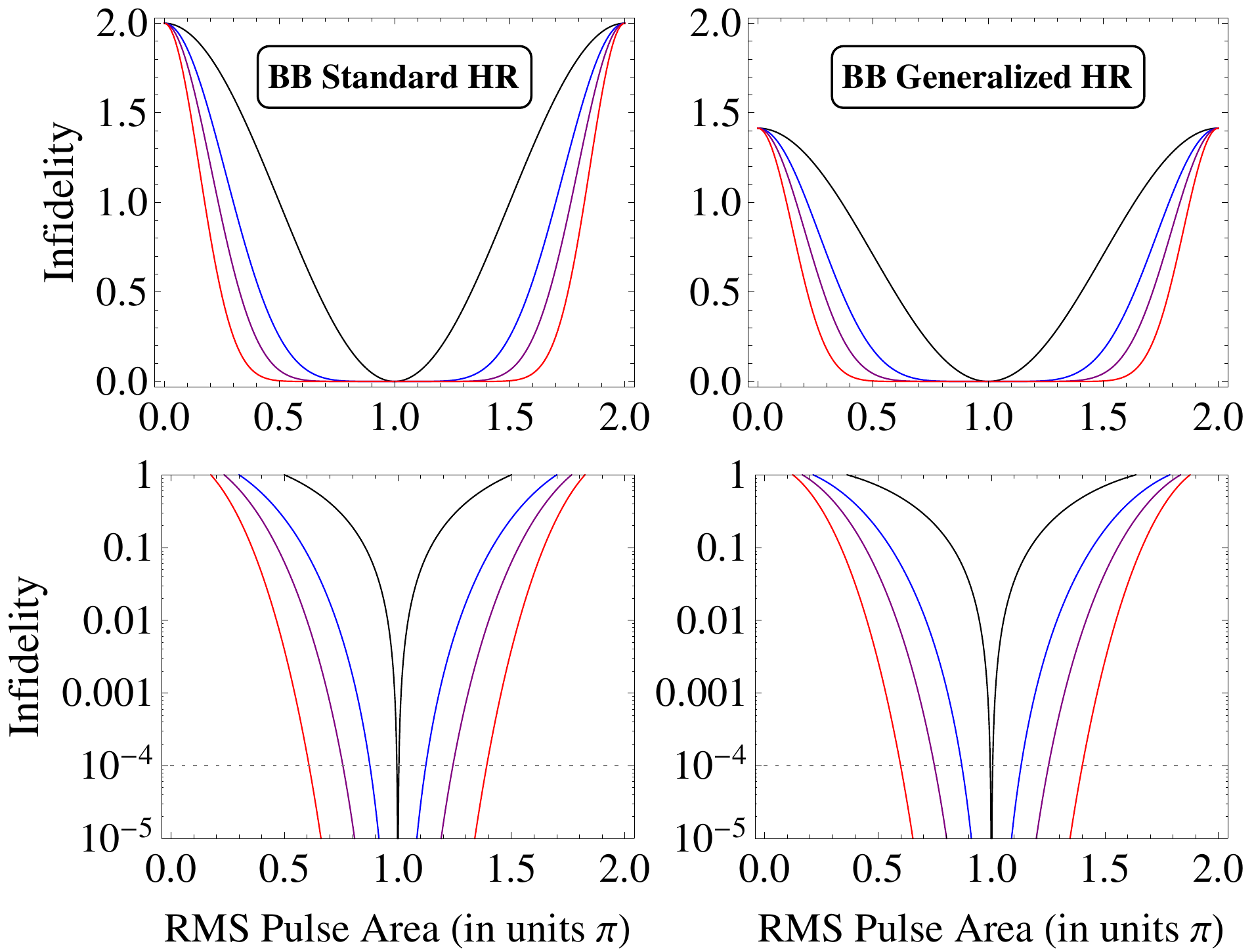}
\caption{Infidelity of the broadband HR as a function of the rms pulse area, for $n=1,3,5,9$ (from inside to outside). 
The left frames refer to the standard HR ($\varphi=\pi$) and the right frames to the generalized HR with $\varphi=\pi/2$. 
Lower frames show the same infidelity as upper frames, but in a logarithmic scale.}
\label{BBHR}
\end{figure}

As seen in the previous section, in order to create a generalized HR, we need to set the value of the Cayley-Klein parameter to $a=\e^{\i\varphi}$. This corresponds to the creation of a phase gate, which, as recently demonstrated \cite{PhaseGate}, can be constructed by using CPs.
As a simplest example, we will first examine the broadband composite HR, which is robust against variations in the pulse area.
Such an HR can be produced by a sequence of two broadband CPs. These pulses have been studied and demonstrated in the literature \cite{Levitt86}. Here, we use the symmetric resonant pulses, derived in \cite{Torosov2011PRA}, with phases given by the formula
\be\label{BBphases}
\phi_k =k(k-1)\frac{\pi}{n}\quad (k=1,2,\ldots ,n),
\ee
and the phases of the second CP are given by Eq. \eqref{phases second CP} for $\alpha=2\varphi$. Explicitly, the phases of the first few CP pulses are (modulo $2\pi$)
\bse
\begin{align}
& \left( 0,\tfrac23,0 \right)\pi, \\
& \left( 0,\tfrac25,\tfrac65,\tfrac25,0 \right)\pi, \\
& \left( 0,\tfrac27,\tfrac67,\tfrac{12}7,\tfrac67,\tfrac27,0 \right)\pi, \\
& \left( 0,\tfrac29,\tfrac23,\tfrac43,\tfrac29,\tfrac43,\tfrac23,\tfrac29,0 \right)\pi.
\end{align}
\ese

\begin{figure}[t]
\includegraphics[width=8cm]{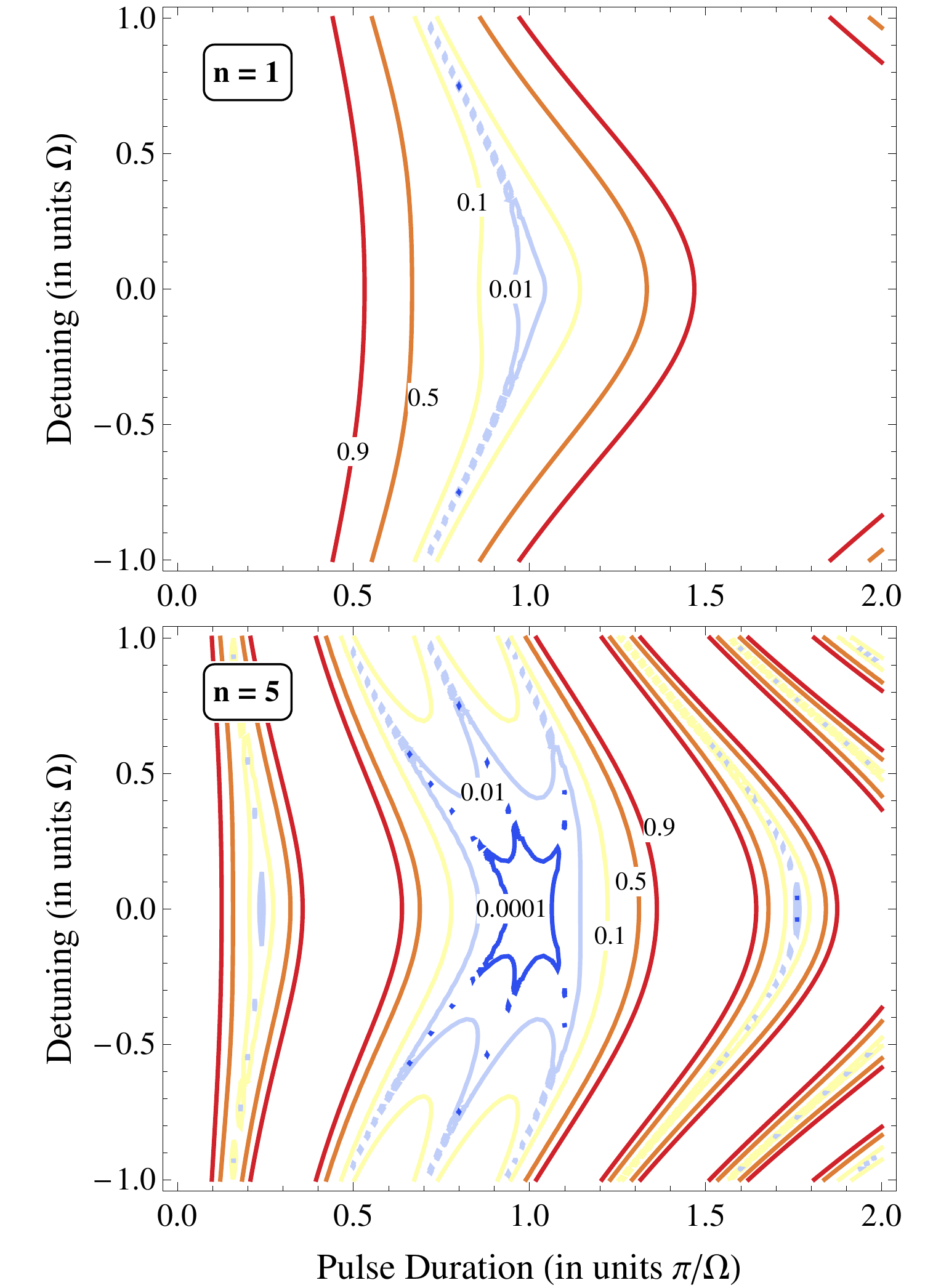}
\caption{Infidelity of the universal standard HR as a function of the pulse duration and the detuning. 
The pulse shape is rectangular and the composite phases in the bottom frame are $\phi_k =(0,11,2,11,0)\pi/6$.}
\label{Uni5}
\end{figure}

As already noted, we assume that all the fields, which couple the $N$ ground states to the excited one, are produced by a single source. 
This means that a systematic error in this source will translate into the same systematic error of all the couplings, which allows the treatment of our system in the same way as in the case of the two-state phase gate.
To test the performance of the composite HRs, we define the infidelity as the Frobenius norm of the distance between the actual operation $\M^{\prime}$ and the desired HR $\M(v,\varphi)$,
\be
F=\sqrt{\sum_{jk}^{N}|M^{\prime}_{jk}-M_{jk}|^2} .
\ee
It can be shown that the infidelity does not depend on the HR dimension $N$ and on the target vector $\ket{v}$, and in the case of BB HRs is given by the simple analytical formula
\be
F=2\sin\varphi/2\cos^{2n}A/2 ,
\ee
where $A=\chi\int_{t_i}^{t_f}f(t)dt$ is the rms pulse area. 
In Fig.~\ref{BBHR} we plot the infidelity of the BB composite standard and generalized HR, for a phase of $\varphi=\pi/2$.
One can see that by increasing the number of pulses, the robustness of the operation also increases.

Continuing the analogy between composite phase gates and composite HRs, one can also build adiabatic composite HR, detuning-compensated HRs, etc.
 
Particularly interesting are the universal composite HRs, which allow compensation of systematic errors in \emph{any} parameter of the field.
This is achieved by using the recently developed universal composite pulses \cite{Genov2014PRL}. The phases of the universal CPs, for $n=3,5,7$, are 
\bse
\begin{align}
& \left( 0,\tfrac12,0 \right)\pi, \\
& \left( 0, \tfrac56, \tfrac13, \tfrac56, 0 \right)\pi, \\
& \left( 0, \tfrac{11}6, \tfrac13, \tfrac{11}6, 0 \right)\pi, \\
& \left( 0, \tfrac{11}{12}, \tfrac{5}{6}, \tfrac{17}{12},\tfrac{5}{6},\tfrac{11}{12},0 \right)\pi, \\
& \left( 0, \tfrac{23}{12}, \tfrac{5}{6}, \tfrac{5}{12},\tfrac{5}{6},\tfrac{23}{12},0 \right)\pi,
\end{align}
\ese
where for $n=5$ and $n=7$ we have two different universal CP solutions.
A contour plot of the infidelity for the universal composite HR is shown on Fig. \ref{Uni5}.


\section{Conclusion}

We have proposed a fault-tolerant implementation of the Householder reflection operator.
The implementation uses the concept of composite pulses, which is a well-developed technology in current experiments, and hence the proposed method is amenable to a relatively simple physical realization.
The proposed implementation requires a good control of the relative phases of the pulses in the composite sequence and the ratios of the couplings in the $N$-pod system.
In return, it is highly accurate and highly insensitive to errors in the other experimental parameters, which makes it suitable for high-fidelity quantum control and quantum information processing.
\acknowledgments

BTT and NVV acknowledge financial support by the EC Seventh Framework Programme under Grant Agreement No. 270843 (iQIT), Bulgarian NSF grants DRila-01/4 and DMU- 03/103. EK and BTT acknowledge financial support by an SUTD Start-Up Research Grant, Project No. SRG EPD 2012 029 and an SUTD-MIT International Design Centre (IDC) Grant, Project No. IDG 31300102.



\begin{thebibliography}{99}
\bibitem{note} We have chosen the MS basis such that the rms Rabi frequency is real. However, a common phase shift in the couplings $\Omega_k$ would still translate as a phase shift in the rms Rabi frequency $\Omega$.
\end{thebibliography}
\end{document}